\documentclass[11pt,A4,twoside]{article}
\usepackage{amssymb}
\usepackage{amsmath}
\usepackage{amsfonts}
\usepackage{amsbsy}
\usepackage{natbib}
\usepackage{eucal}
\usepackage{graphicx}
\usepackage{titlesec}
\usepackage[english]{babel}
\usepackage{times}
\usepackage{mathptmx}
\usepackage{float}
\usepackage[font=small]{caption}
\usepackage[colorlinks,citecolor=blue,urlcolor=blue]{hyperref}
\usepackage{authblk}
\usepackage{amsthm}
\usepackage{titling}

\usepackage{setspace}
\DeclareSymbolFont{txgreek}{OML}{cmr}{m}{it}

\renewcommand{\abstract}[1]{{\small\noindent
\hrulefill\par \vspace*{0.1cm}\noindent{\small\bf\sffamily
{Abstract}}\parindent=0pt\par\noindent\vspace{-0.1cm}\noindent\hrulefill\par\vspace*{0.5\baselineskip}\hspace*{0cm}\renewcommand{\baselinestretch}{1.1}\sffamily{#1}\par\vspace*{-0.1cm}\noindent\hrulefill}}

\def\paragraf{\fontsize{9}{10pt}\fontfamily{phv}\fontshape{it}\selectfont}
\titleformat{\paragraph}
{\titlerule[0pt]
\vspace{0cm}%
\paragraf}
{\theparagraph}{.5em}{\vspace*{0\baselineskip}}

\def\titol{\fontsize{12.045}{12pt}\fontfamily{phv}\fontseries{b}\selectfont}
\titleformat{\section}
{\titlerule[0pt]
\vspace{0.2cm}%
\titol}
{\thesection.}{.5em}{\vspace*{0\baselineskip}}

\def\titolp{\fontsize{11.045}{11pt}\fontfamily{phv}\fontseries{b}\fontshape{it}\selectfont}
\titleformat{\subsection}
{\titlerule[0pt]\bigskip
\vspace{-0.4cm}%
\titolp}
{\thesubsection.}{.5em}{\vspace*{0\baselineskip}}

\def\titolpp{\fontsize{10.045}{10pt}\fontfamily{phv}\fontshape{it}\selectfont}
\titleformat{\subsubsection}
{\titlerule[0pt]
\vspace{-0.4cm}%
\titolpp}
{\thesubsubsection.}{.5em}{\vspace*{0\baselineskip}}

    \predate{}
    \date{}
    \postdate{}

\title{A Bayesian - Deep Learning model for estimating Covid-19 evolution in Spain.}

\thanksmarkseries{arabic}
\author{Stefano Cabras \thanks{Department of Statistics, Universidad Carlos III de Madrid (Spain), stefano.cabras@uc3m.es }}

\def\headers#1{\fontsize{8.5}{10}\centering\sffamily\itshape{#1}}
\def\page#1{\fontsize{8.5}{10}\sffamily{#1}}
\usepackage{fancyhdr}

\begin{document}
\maketitle

\thispagestyle{empty}
\renewcommand{\headrulewidth}{0truecm}
\pagestyle{fancy}
\rhead[\headers{Bayesian-DL for Covid-19 in Spain}]{\page{\thepage}}
\lhead[\page{\thepage}]{\headers{Stefano Cabras}}
 \lfoot{} \rfoot{}
\cfoot{}

\abstract{This work proposes a semi-parametric approach to estimate Covid-19 (SARS-CoV-2) evolution in Spain. Considering the sequences of 14 days cumulative incidence of all Spanish regions, it combines modern Deep Learning (DL) techniques for analyzing sequences with the usual Bayesian Poisson-Gamma model for counts. DL model provides a suitable description of observed sequences but no reliable uncertainty quantification around it can be obtained. To overcome this we use the prediction from DL as an expert elicitation of the expected number of counts along with their uncertainty and thus obtaining the posterior predictive distribution of counts in an orthodox Bayesian analysis using the well known Poisson-Gamma model. The overall resulting model allows us to either predict the future evolution of the sequences on all regions, as well as, estimating the consequences of eventual scenarios.}

\paragraph{MSC: 62P10}

\paragraph{Keywords: Applied Bayesian methods, Deep Learning, Multivariate Time Series, LSTM, SARS-CoV-2 (Covid-19)}

\renewcommand{\baselinestretch}{1.2}
\bigskip

\section{Introduction}\label{secintro}

Understand and predict the evolution of Covid-19 (SARS-CoV-2) diffusion has become of primary importance in the actual Spanish society. The study of a disease spread is an old topic in statistical epidemiology and the actual disease is not an exception. Much of epidemiology monitoring models rely on the well-known SIR model in which the basic reproduction rate is an indicator of infection evolution. The reproduction rate is a function of the number of cases and these are our primary quantity of interest from which many others and not only the reproduction rate, can be derived. The SIR model in its basic version assumes that the model parameters are constant along with time and space although variations along time are currently under development \cite{chen2020time}. However, since the beginning of the application of spatial-temporal statistical analysis, modelling has evolved incorporating into the epidemiology of more sophisticated models. These are derived from statistical applications in different fields. For instance, in time-series analysis Covid-19 evolution is viewed as a sequence of counts (of cases, deaths, etc..), and autoregressive and moving average models can be used \citep{agosto2016modeling,agosto2020poisson}. From a spatial analysis, we consider the evolution of Covid-19 at the areal level (in contrast to the point data approach) in a specified time domain. We feel that Bayesian modelling, more than just another way to build models with complicated structures, is necessary as it takes into account the uncertainty that data, especially noisy data, bear to any statement on Covid-19 evolution. In this context, Bayesian models derived from time series analysis with random spatial effects have been proposed in \cite{paul2011predictive} for epidemic monitoring and already applied to Covid-19 diffusion in Italy \cite{giuliani2020modelling} (a generalization of \cite{agosto2020poisson}). The problem with the usual Bayesian spatial-temporal model is that they assume specific parametric forms for evolution along time (e.g.,  ARMA process) and for spread among areas mostly given by a neighbourhood matrix plugged in a Conditional Autoregressive (CAR) Model which assumes that neighbourhoods are defined beforehand, in contrast, to be adaptive on the observed data.

It is well known that the disease diffusion process is not linear and while surrounding areas are a good approximation at a local level of disease spread, these may be not optimal at a larger scale. Considering the specific case of Spain, we can for instance state that although Cataluña and Canary Islands are not surrounding areas of Madrid, these are connected to the Madrid region by high-speed trains and flights. This is maybe enough to put in question the usefulness of the concept of the neighbourhood surrounding areas assumption defined beforehand and encoded in the neighbourhood matrix.

We account for interesting and still not modelled dynamics: the actual number of cases may prelude to future increments and it can be explained by the past number of cases into other regions. The question is: does it make sense to read the sequence of counts from the past to future (as usual in time series), or also should we read them in the opposite (retrospective) way? We do not address this question, but instead, we analyze counts in both directions. This is precisely what we try to do in reading a text: a new word sometimes can be understood reading forward rather than just looking backwards.

To complicate things even more, we have to take into account the fact that the data collecting process is far from being regular. First of all, in Spain, there is no centralized official unit dedicated to collect Covid-19 data assuring data consistency. There only exists, at the day of writing, groups of academics that spontaneously collect data and share it (see, for instance, \url{https://github.com/datadista/datasets/tree/master/COVID\%2019}, code.montera34.com), while in other countries as in Italy there exists an official repository of Italian civil guard (\url{https://github.com/pcm-dpc/COVID-19}). However, even with this official data are still questioned for being useful in analyzing Covid-19 in Italy \cite{bartoszek2020official}. In this paper, we will only use the version of the counts provided by the Instituto Carlos III de Salud (\url{www.isciii.es}). Such noise in the data requires robust statistical methods.

In practice, Covid-19 spread is the result of complex dynamics along with time and space that can either be modelled by good experts or try to estimate from the available noisy data. The latter approach is chosen as modern machine learning techniques are here used to create such a super expert (from the data of course) that will elicit priors on a Bayesian model for counts. The discrepancy between observations and experts is properly taken into account by Bayes theorem which finally allows predicting counts along with the corresponding uncertainty.

The rest of the paper is organized as follows: Section \ref{secmodel} discusses the two-step analysis whose results are reported in Section \ref{secappcovid}. With the proposed approach we are also able to forecast the eventual scenarios that are presented in Section \ref{secscenarios}. Section \ref{secconclusion} reports possible generalizations of the approach, which are beyond the scope of the paper, but that can be quickly implemented using the code in the GitHub project: \url{https://github.com/scabras/covid19-bayes-dl}.

\section{Model}\label{secmodel}
Let $Y_{ts}\in 0,1,2,\ldots$ be the random variable of interest representing the 14 days cumulative incidence per 100.000 inhabitants  in region $s=1,\ldots,S=19$ at day $t=1,\ldots,T$. The aim of the paper is to estimate
\begin{equation}
P|\mathcal{D}=Pr(Y_{ts}=y|\mathcal{F}_{t-1})|\mathcal{D}, \label{eqpp}
\end{equation}
where $\mathcal{F}_{t-1}$ represents the process filtration up to the day $t$. The Bayesian approach is needed to properly account for conditioning on observed data $\mathcal{D}$, while the non-parametric approach implemented in the Deep Learning (DL) model enters as $\mathcal{F}_{t-1}$ is the evolution of all sequences, that is overall possible $s$ and up to the day $t-1$. Estimating conditionally on such $\mathcal{F}_{t-1}$ it would be possible even with parametric models but at the cost of requiring strong expertise on fixing, beforehand, the parametric form of $P$ by including how to relate past and future Covid-19 evolution.

To avoid assumptions derived from such an unavailable expert, we only fix the family of $P$ being the Negative Binomial, derived from the usual Poisson likelihood and a Gamma prior on the Poisson mean that will be specified below. Hence,

$$Y_{ts}|\eta_{ts} \sim  Poisson(\exp(\eta_{ts})),$$

where $E(Y_{ts})=\eta_{ts}$.

For estimating $\eta_{ts}$ we first project all observations up to $t-1$ into a point guess $\widehat{y}_{ts}$ using the later specified biderectional Long Short Term Memory (LSTM) model having as input all sequences up to day $t-1$. This is illustrated in Section \ref{seclstm}. Secondly, we derive the posterior distribution of $\eta_{ts}$ by assuming {\it a priori}, $E(\eta_{ts})=\widehat{y}_{ts}$ and variance $Var(\eta_{ts})=E((\widehat{y}_{ts}-Y_{ts})^2)$, where the latter is obtained conditionally to $s$ and to the {\em delay} in predicting, e.g. eliciting the mean 7 days ahead bear more uncertainty than eliciting one day ahead. Bayes theorem finally allows to obtain \eqref{eqpp} as illustrated in Section \ref{secbayes}.

\subsection{Long Short Term Memory (LSTM)}\label{seclstm}

A deep learning (DL) model is a neural network with many layers of neurons \cite{schmidhuber2015deep}, it is an algorithmic approach rather than probabilistic, see \cite{breiman2001statistical} for the merits of both approaches. Each neuron is a deterministic function such that a neuron of a neuron is a function of a function along with an associated vector of weights $\mathbf{w}=(w_1,\ldots,w_k)$. Essentially, for a generic response variable $Y_i$ of the $i$th statistical unit and a corresponding predictor $X_i$ we have to estimate
\begin{equation}
    Y_i = w_1f_1(w_2f_2(...(w_kf_k(X_i)))).\label{eqdlrep}
\end{equation}
The larger $k$ is, the "deeper" is the network. With many stacked layers of neurons all connected (a.k.a. dense layers), it is possible to capture high nonlinearities and interactions among variables. The approach to model estimation underpinned by a DL model is that of a compositional function in contrast to that of additive function underpinned by the usual regression techniques including the most modern one (e.g. Smoothing Splines, Non-parametric regression, etc...), as $Y_i = w_1f_1(X_i)+w_2f_2(X_i)+...+w_kf_k(X_i)$. A throughout review of DL is beyond the scope of this paper and can be found, for instance at \cite{schmidhuber2015deep}.

When $f(X)$ functions are linear in its argument, the DL model can be also interpreted as a maximum \textit{a posteriori} estimation of $\Pr(Y|X,Data)$ for Gaussian process priors \cite{polson2017deep}. However, despite this and because of its complexity it cannot be evaluated the whole distribution $\Pr(Y|X,Data)$, but only its mode.

Fitting a DL consists of estimating the vectors of weights $\mathbf{w}$. The estimation requires evaluating a multidimensional gradient which is not possible to be performed jointly for all observations, because of its dimensionality and complexity. Recalling that the derivative of a composite function is defined as the product of the derivative of inner functions (i.e. the well-known chain rule $(f\circ g)' = (f'\circ g)\cdot g'$), this is implemented for purposes of computational feasibility as a tensor product which in turn is implemented in the hardware (not software) of Graphical Processing Unit (GPU) device. A GPU can deal in parallel with thousands of treads at a relatively lower cost than the usual eight treads of a normal CPU. That is what makes the DL approach to data analysis popular in recent years as it can be run also on a mobile phone. Such tensor products, along with the entirely DL machinery, are evaluated for batches of observations (in this case sequences of counts) and it is implemented, for instance, in the open-source software known as Google Tensor Flow \cite{tensorflow2015-whitepaper}.

Most of the DL models are suitable for independent observations in which batches of observations can be drawn at random from the sample and then used to estimate $\mathbf{w}$. Such models cannot be used here as observations are sequences and thus not independent. For these purposes, we have to resort to specific DL models as those belonging to the class of recurrent neural networks (RNN) \cite{lim2020time}. Fundamentals of RNN and Long Short Term Memory (LSTM) models (a specific instance of RNN) can be found here \cite{lstm-formalism}. Unfortunately, this paper needs some translation to a statistical audience, but basically, an RNN acts similarly to a Hidden Markov Model (HMM) or more specifically to a Dynamic linear model \citep{harrisonwest} for observed sequences $Y_1,\ldots, Y_{t}$. They differ from usual HMM in that they do not manage conditional probabilities, but signals which can be viewed as point guess of observed and hidden states represented by the above mentioned deterministic nodes of the DL network. The modeling approach instead of being justified on Markovian processes is derived from posing an additive deterministic model on the evolution equation evaluated at time $t^*$, $\frac{dY(t)}{dt}\mid_ {t=t^*}$ with the same structure as in \eqref{eqdlrep} in which the evolution of sequence at time $t^*$ $\frac{dY(t)}{dt}\mid_ {t=t^*}$ is the response and $\left(\frac{dY(t)}{dt}\mid_ {t=t^*-1},\ldots,\frac{dY(t)}{dt}\mid_ {t=t^*-k}\right)$ are the predictors. In this case, $k$ has the meaning of lag in time series analysis and also the hidden layers in this model architecture.

RNN with says $k$ hidden layers implies that the evolution at time $t^*$ is a nonlinear function of $k$ past evolution equations. The problem is that if the observations were not informative for estimating such a complex function with the corresponding $\mathbf{w}$, then the gradient, would vanish and thus $\mathbf{w}$ cannot be updated. This prevents the train of RNN architecture. To avoid the vanishing gradient problem, the LSTM model introduces an adjustment \citep{lstm-formalism} of the gradient which avoids it being exactly zero. This allows us to estimate short effects (terms of the evolution equation near to $t^*$) as well as long-term effects ($t$ far from $t^*$).

The structure of the network can be further complicated by connecting an LSTM network that analyzes sequences in the order $1,\ldots,t-1,t$ with another LSTM network that analyzes sequences in the opposite order $t,t-1,\ldots,1$. These types of architectures are called bidirectional LSTM. They are used for Artificial Intelligence (AI) language understanding and speaking, described, for instance, in \cite{wollmer2010bidirectional} and implemented in Google Tensor Flow.

To account for spatial relations, the output of the Bidirectional LSTM is further connected among the $S$ series, one per region, by a dense layer which is a layer of all connected nodes. This assures that, if there exists a correlation between region dynamics, this is properly accounted into the model. Note that the model can look at correlation dynamics, not just correlations among time series. That is, if in one region cases are increasing or decreasing in some way, this could be correlated with the dynamics of cases in other regions.

Finally, if we let $Y$ being as a vector of $S$ time series each one with length $T$ and $X$ the corresponding vector of past (future, depending on the direction) values of $Y$ each of length 14 days. We use a window time of 14 days to predict the next $d \geq 1$ days ahead. Once $\mathbf{w}$ have been estimated (i.e., the network has been trained), we end up in having the guess, $\widehat{y}_{ts}$, of the mode of $Pr(Y_{ts}|\mathcal{F}_{t-1})$ in \eqref{eqpp} as the bidirectional LSTM model jointly considers the past evolution of the Covid-19 in Spain.

The guess $\widehat{y}_{ts}$ represent a suitable projection of all sequences evolution into the space of sequence guesses, just as a sample mean is used to project observed quantities or used in eliciting priors on historical data. If we want to recover the Bayesian analysis, it can be considered as elicitation of the expected number of counts at time $t$ and region $s$. Also in an empirical Bayes point of view it can be considered as a point estimation of the prior hyper-parameter $E(Y_{ts})$ by setting $\widehat{E(Y_{ts})}=\eta_{ts}$.

It is worth noting that predictions of a sequence of unobserved days have been performed autoregressively. That is, the prediction of the day $d=1$ is used as input for predicting the next day $d=2$ and so on. We consider up to $d=7$ days ahead.

It remains is to assess the uncertainty around such elicitation. This is possible by looking at the prediction error of the LSTM architecture of $d\geq 1$ days ahead predictions. This is done by evaluating the variance of the prediction along with all the trained sequences.  Note that this is calculated as the difference between the observed count (not included in the trained sequence up to that day) and the predicted count by the network using observations $d$ days before the predicted day. There are

This will also provide a reliable assessment of the LSTM predictions.

\subsection{Poisson-Gamma model}\label{secbayes}

The aim of this second and last step in the analysis of the cumulative incidence $Y_{ts}$ is to derive the uncertainty for counts at time $t$, region $s$ given the elicited mean $\widehat{y}_{ts}$ and the variance  $Var(\widehat{y}_{ts})$ obtained in the previous step described in the above in Section \ref{seclstm}. The model is very simple: for each count,

\begin{eqnarray}
    Y_{ts}|\eta_{ts} & \sim & Poisson(\eta_{ts}),\quad \mbox{(Likelihood)} \label{eqlik}\\
    \eta_{ts}|\mathcal{F}_{t-1} & \sim & Gamma(a,b),\quad \mbox{(Prior)}\label{eqprior}\\
    E(\eta_{ts})& = & ab=\widehat{y}_{ts},\quad \mbox{(Prior mean)} \label{eqpriormean} \\
    Var(\eta_{ts})& = & ab^2 = Var(\widehat{y}_{ts}),\quad \mbox{(Prior variance) \label{eqpriorvar}}
 \end{eqnarray}

and the prior information is that of the LSTM expert with the reliability given by its average prediction error.

The predictive posterior \eqref{eqpp} at an observed count $y_{ts}$ is a Negative Binomial:

\begin{equation}
    \Pr(Y_{ts}=y|\mathcal{D})=\frac{\Gamma\left(a+y_{ts}+y\right)}{\Gamma(y+1) \Gamma\left(a+y_{ts}\right)}\left(\frac{b+1}{b+2}\right)^{a+y_{ts}}\left(\frac{1}{b+2}\right)^y \label{eqpredpost}
\end{equation}

while the prediction of an unobserved day is obtained by using the corresponding (informative) prior predictive distribution:

\begin{equation}
    \Pr(Y_{ts}=y|\mathcal{D})=\frac{\Gamma\left(a+y\right)}{\Gamma(y+1) \Gamma\left(a\right)}\left(\frac{b}{b+1}\right)^a\left(\frac{1}{b+1}\right)^y \label{eqpriorpred}
\end{equation}

For a given $a$ and $b$ the predicted variance of \eqref{eqpriorpred} is greater than that of \eqref{eqpredpost}

All other quantities of interest such as cumulative counts, reproduction rate, and aggregation at higher territorial levels, can be obtained by simulating from \eqref{eqpredpost} and \eqref{eqprior}.

\section{Covid-19 modelling}\label{secappcovid}

In this section, we apply the above model to the problem of estimating Covid-19 evolution in Spanish regions. The analysis is implemented in the following GitHub project: \url{https://github.com/scabras/covid19-bayes-dl}, where code and results can be browsed at \url{https://scabras.github.io/covid19-bayes-dl/} and are updated with newer observations than those considered in this work. Results here refer to data available up to February 18$^th$, 2021.

\subsection{Data}\label{secdata}

Data comes from Instituto Carlos III de Madrid in Madrid and they can be downloaded in CSV format at \url{https://cnecovid.isciii.es/covid19/resources/casos_diag_ccaadecl.csv}. These are assumed as the most reliable data for Spain available at the moment. From the database, we used the total number of reported cases obtained as the sum of the results of the different types of tests. Such a number is not updated regularly, especially at the beginning of the spread, during weekends and also past days may be updated in the future as more information arrives at Instituto Carlos III de Madrid. Hence, we considered the cumulative incidence over 14 days which leads to more reliable information about the dynamic of the pandemic. That is for the number of residents in region $s$, $Pop_s$, the sum over 14 days leading to $Y_{ts} \times Pop_s/100000$ compensates all under sampling effects mentioned above.

\begin{figure}[H]
    \begin{center}
        \includegraphics[width=10cm]{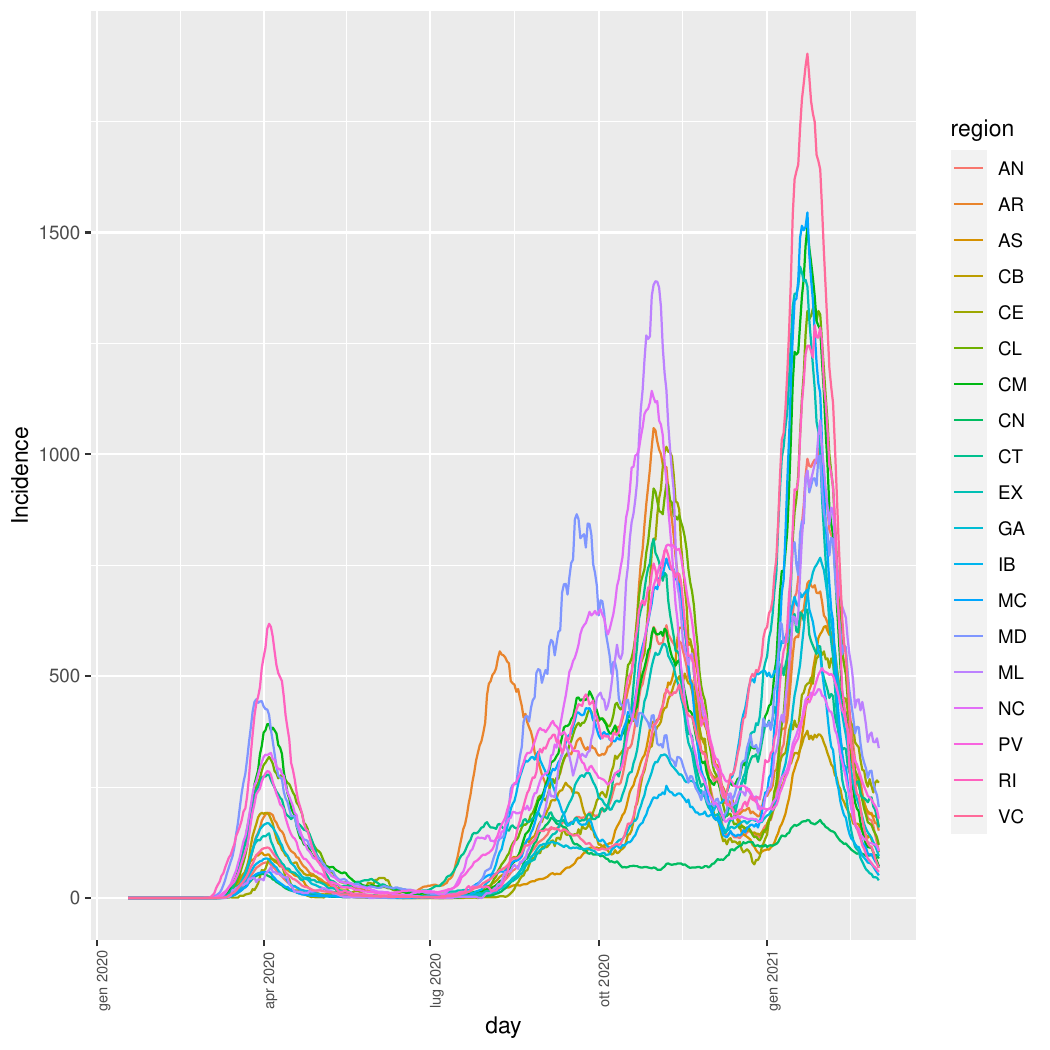}
        \caption{Observed cumulative incidence at 14 days for the 19 Spanish regions. The period starts January 1$^st$ 2020 and end the 18$^th$ of February 2021 for a total number of $T=415$ days by $S=19$ regions.}\label{fig.obsinc}
    \end{center}
\end{figure}

Figure \ref{fig.obsinc} reports the observed incidence, $y_{ts}$ over all regions and all available period. We can appreciate the presence of three waves. The period starts January 1$^st$ 2020 till the 18$^th$ of February 2021 for a total number of $T=415$ days by $S=19$ regions.

Spanish regions are reported into figures through their ISO standards: \url{https://en.wikipedia.org/wiki/ISO_3166-2:ES}.

\subsection{LSTM interpolation}\label{secinterpolation}

The presence of noisy counts reported in Figure \ref{fig.obsinc} justifies the need for using robust models to describe the data. Specifically, we used a bidirectional LSTM model looking back (forward) where the input at each sequence in a batch is $k=14$ days which are the usual two weeks for an asymptomatic case to become symptomatic and hopefully appear in the database as a case. Therefore, at each time $t$ and for the backward direction of the LSTM the process filtration $\mathcal{F_{t-1}}$ refers to the history of the process up to two weeks before. On the opposite, the forward part of the LSTM network is devoted to estimating the evolution of the 2 weeks ahead.

The DL network is made of two stacked layers of neurons: the first is 64 all connected LSTM layers (one for direction) and the second is a layer of $19 \times 7$ (i.e. $d=7$) all connected linear neurons. The model capability is of around 52 thousand weights (nodes are all connected), most of them will be zero as the data are not informative enough to update all original zero weights.

Such architecture models:
\begin{itemize}
    \item in sequence evolution, that is, $Y_{ts}|Y_{t^\prime s}$ for all $t^\prime \neq t$;
    \item between-sequence evolution that is $Y_{ts}|Y_{t^\prime s^\prime}$ for all $t^\prime \neq t$, $s^\prime \neq s$.
\end{itemize}

This translates into saying that for instance, the number of cases in Madrid can be affected by past cases and will affect the future number of cases (in sequence evolution), as well as the cases in Madrid, can be a consequence or can have effects on past and future cases in all other regions (between-sequence evolution).

This is the main contribution of the proposed way of representing data to the actual literature of Covid-19 evolution already mentioned above.

The model has trained 70 steps (epochs) and the training sample is made of batches of 10 random sequences drawn at each optimization step. The training sample per step consists of 10 random sequences of $k+d=21$ subsequent rows of the original data matrix $\mathcal{D}$ of size $T \times S$.

All nodes in the network are linear in their arguments and to further robustify the analysis, the weights are estimated to minimize the prediction error defined as the mean absolute error between the estimated and observed counts.

For the 70 steps Figure \ref{fig.history} (in Appendix) reports the estimation error which decreases with the step meaning that the network is learned from the available sample. Figure \ref{fig.history} can be interpreted also as a lack of overfitting given that the number of parameters is much larger than the number of observed days and the error is still positive, reaching a plateau. This is related to the concept of model capability, which differs from that of model dimension typical in usual regression analysis (without shrinkage methods). In the usual regression, model data are used to impose an estimated value on all coefficients. This is not necessary here and only relevant and consistent information is passed along the nodes of the network. This is why the overfitting is somehow limited and the training error is bounded away from zero.

\begin{figure}[H]
    \begin{center}
        \includegraphics[width=10cm]{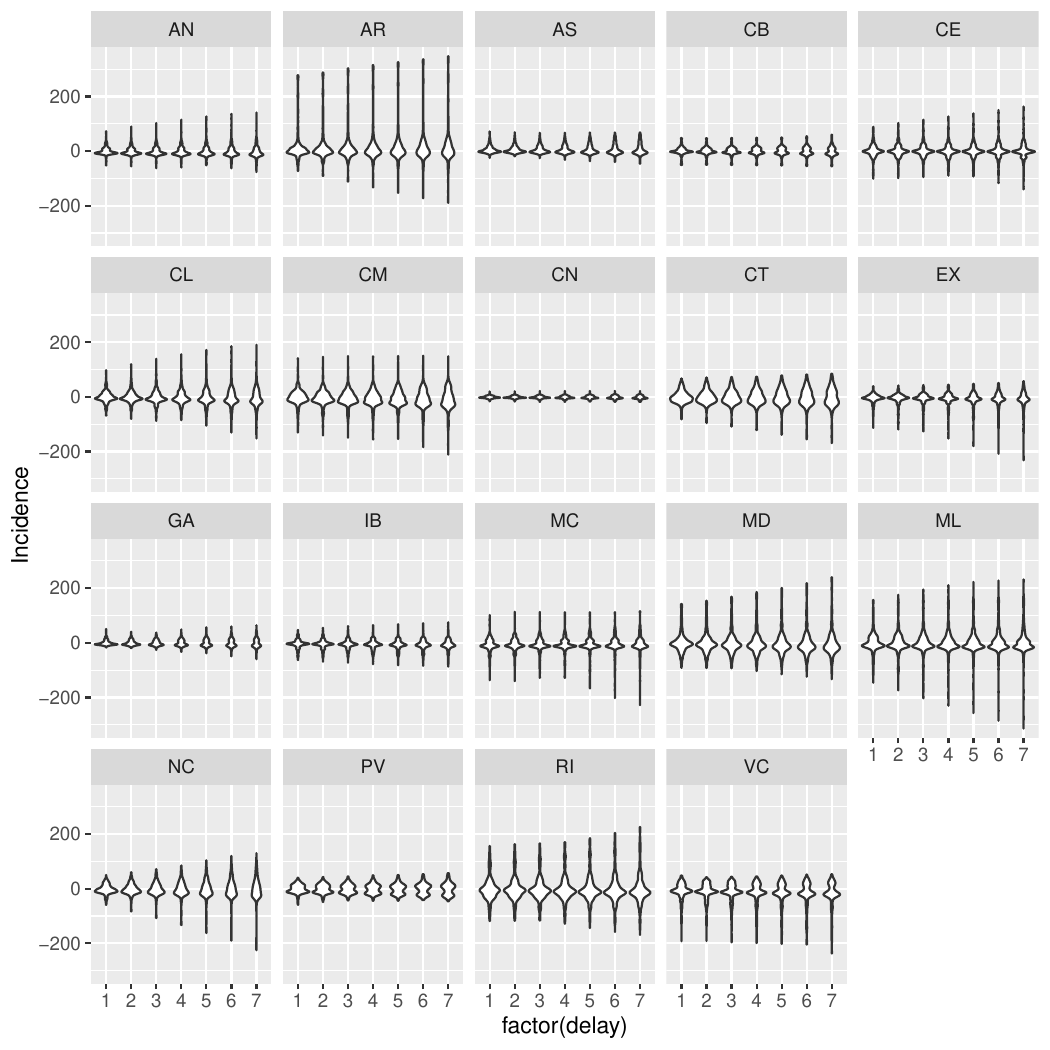}
        \caption{Violin plots of distributions of differences $\widehat{y}_{ts}-y_{ts}$ conditionally to region and the number of days ahead in the prediction, the delay $d$.}\label{fig.delayerrors}
    \end{center}
\end{figure}

Figure \ref{fig.delayerrors} reports the distribution of differences $\widehat{y}_{ts}-y_{ts}$ conditionally to region and the number of days ahead in the prediction, the delay $d$.

\begin{figure}[H]
    \begin{center}
        \includegraphics[width=10cm]{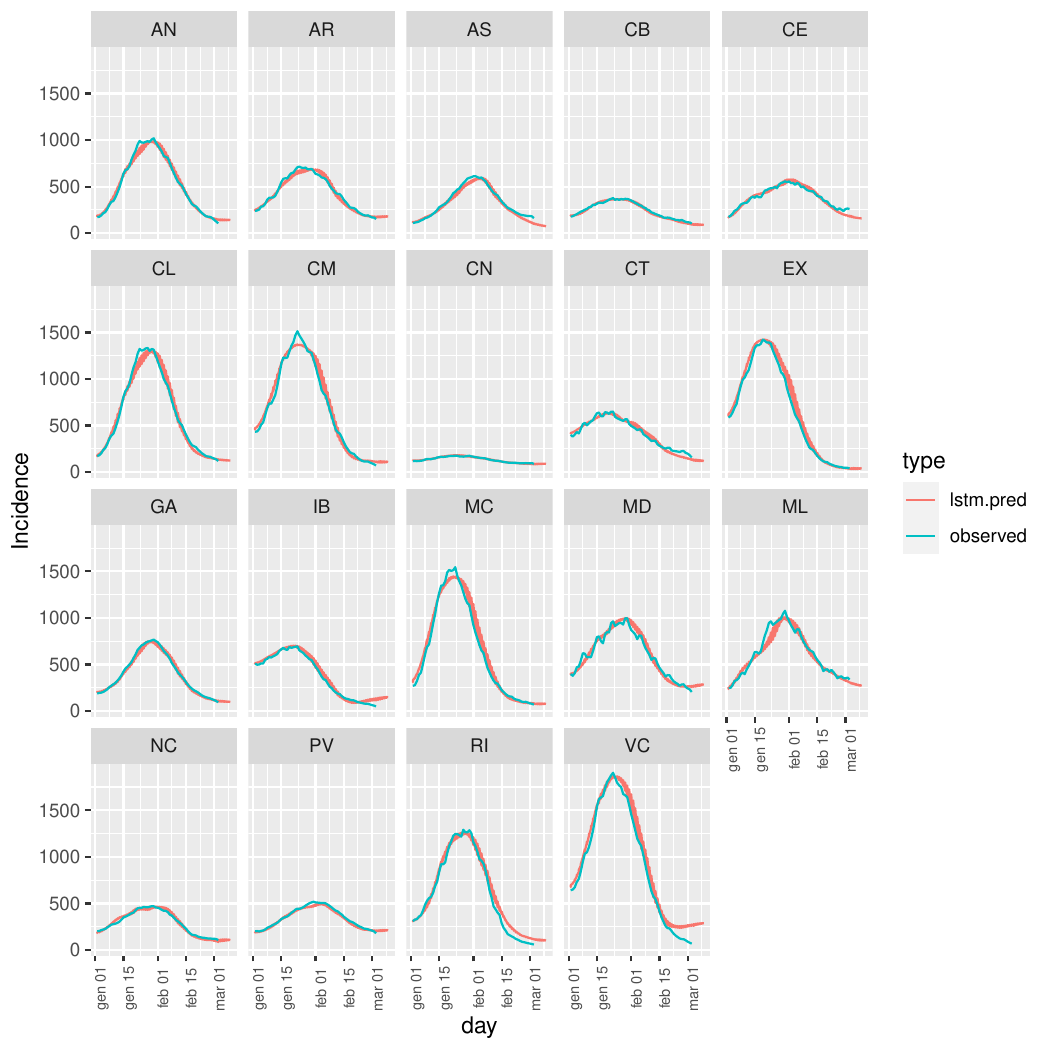}
        \caption{Predicted $\widehat{y}_{ts}$ and observed daily incidence over 2021 only.}\label{fig.lstmpred}
    \end{center}
\end{figure}

The variance of each distribution is $Var(\widehat{y}_{ts})$, used as the prior variance \eqref{eqpriorvar}. The general error is around 0 and predicting with more days ahead leads to greater errors although this is not constant over regions and there are regions whose dynamics is more difficult to be predicted. This is again corroborated in Figure \ref{fig.lstmpred} where we can see that observed $y_{ts}$ and predicted $\widehat{y}_{ts}$ are distinguishable at the end of the third wave.

It seems that the LSTM would not expect the rapid decrease of cases during February 2021 in many regions except, for instance, Asturias (AS), Cantabria (CB) or Canary Islands (CN). Looking at the prediction over the whole period (Figure \ref{fig.lstmpredall} in Appendix), predicted and observed are barely distinguishable suggesting that the LSTM expert is generally reliable for eliciting prior \eqref{eqprior}.

Even if the observed sequences are noisy, the point is that such noise is common in many regions and paradoxically it turns out to be "regular noise", that is, the network can detect the effect of reporting counts during and just after weekends. This could have been not possible to be accounted for using simpler linear regression models like those mentioned in the introduction unless very informative experts were available for assessing interactions among the 19 series of counts.

The LSTM "expert" indicates that the incidence decreasing is slowing down in many regions. Of course, we have no clue how much are probable the above statements to be true and this is addressed in the following Section \ref{secpp}. The additional problem in using directly the output of a DL model regards the fact that it is very difficult to understand why are that decrease/increase and this casts doubt on the direct use of such an output as a final prediction. Although this, it remains the usually observed sampling argument: why a model that did well in the past should fail in the future? Is this due to overfitting? This is the question that also justifies the subsequent Bayesian analysis to at least put probability evaluations to the two statements of the question: 1) how well did in the past? and 2) how much to trust the predicted future? For the second question, we already estimated the prediction errors on $d$ days ahead and given that the number of time points is large, then $Var(\widehat{y}_{ts})$ estimated from Figure \ref{fig.delayerrors} is now a reliable estimation of the model goodness of fit because the predicted day is not into the $k$ past days used to predict it.

\subsection{Bayesian predictive analysis}\label{secpp}

Results on Figure \ref{fig.lstmpred} and the above LSTM model do not convey a proper evaluation of the uncertainty around the estimation of one day ahead evolution. For this purpose, we calculate the posterior predictive distribution of $Y_{ts}$ according to \eqref{eqpredpost} and \eqref{eqpriorpred}. The key is to use the above AI expert which elicits priors on Poisson intensity according to \eqref{eqprior}. Such an expert can be trusted based on the pros and cons described in Section \ref{secinterpolation} and also we have a reliability measure of the expert for $d$ days ahead prediction. In particular, for already observed days, we consider $d=1$ as the most reliable elicitation and $d>2$ for the rest of unobserved days.

\begin{figure}[H]
    \begin{center}
        \includegraphics[width=10cm]{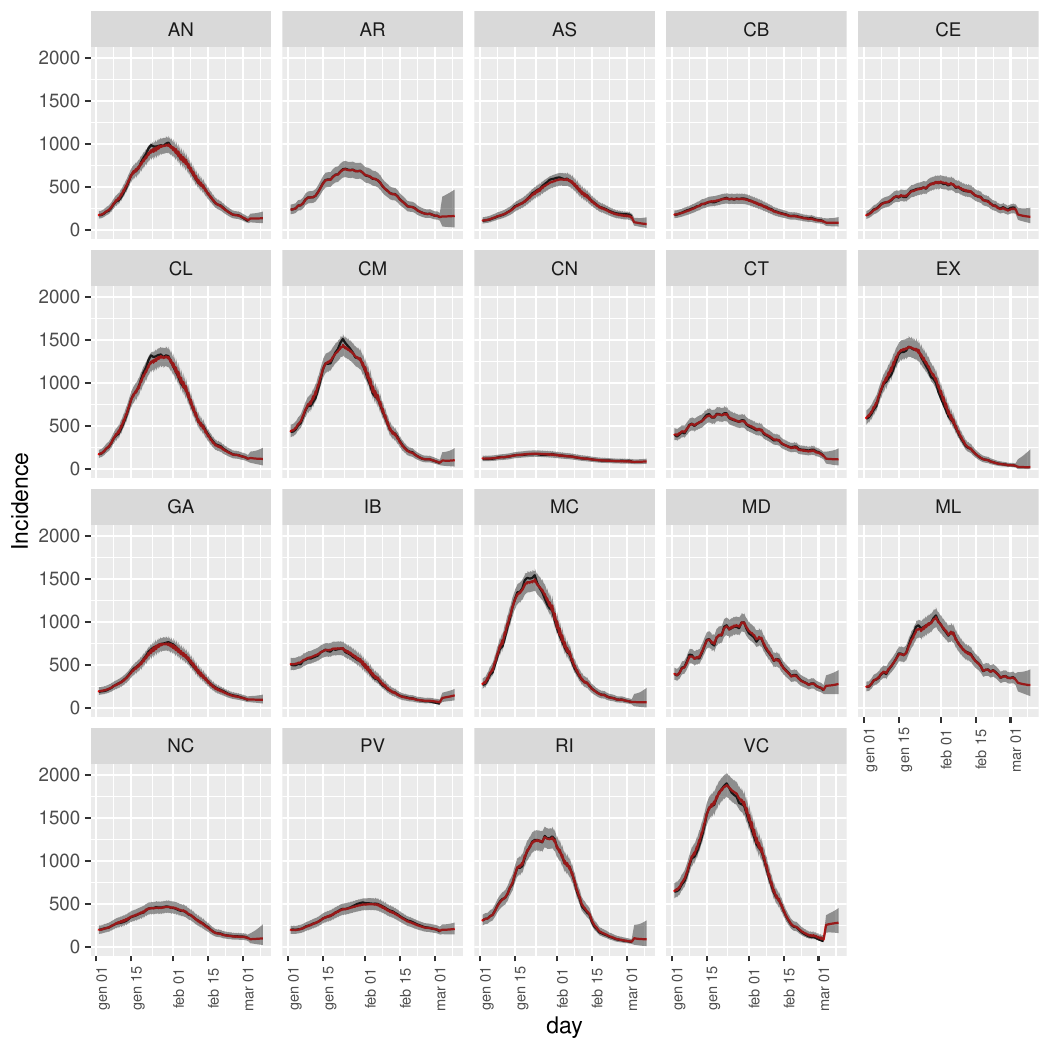}
        \caption{Daily incidence for year 2021. Posterior predicted distribution summarized by the median (red) and 99\% credible intervals (shaded grey area) for observed days (black) This is the estimation of \eqref{eqpp} for observed and future predictions starting from $d=1$ day ahead up to $d=7$ days ahead.}\label{fig.bayespred}
    \end{center}
\end{figure}

The result is the prediction along with its uncertainty, which is the estimation of \eqref{eqpp}. This is the base result of all modelling process and in Figure \ref{fig.bayespred} we report the observed and predicted counts along with 99\% equal tail credible intervals (CI).

From Figure \ref{fig.bayespred} we can see that accounting for the uncertainty, the observed values are compatible with 99\% CI and. This uncertainty increases for unobserved days.

\begin{figure}[H]
    \begin{center}
        \includegraphics[width=10cm]{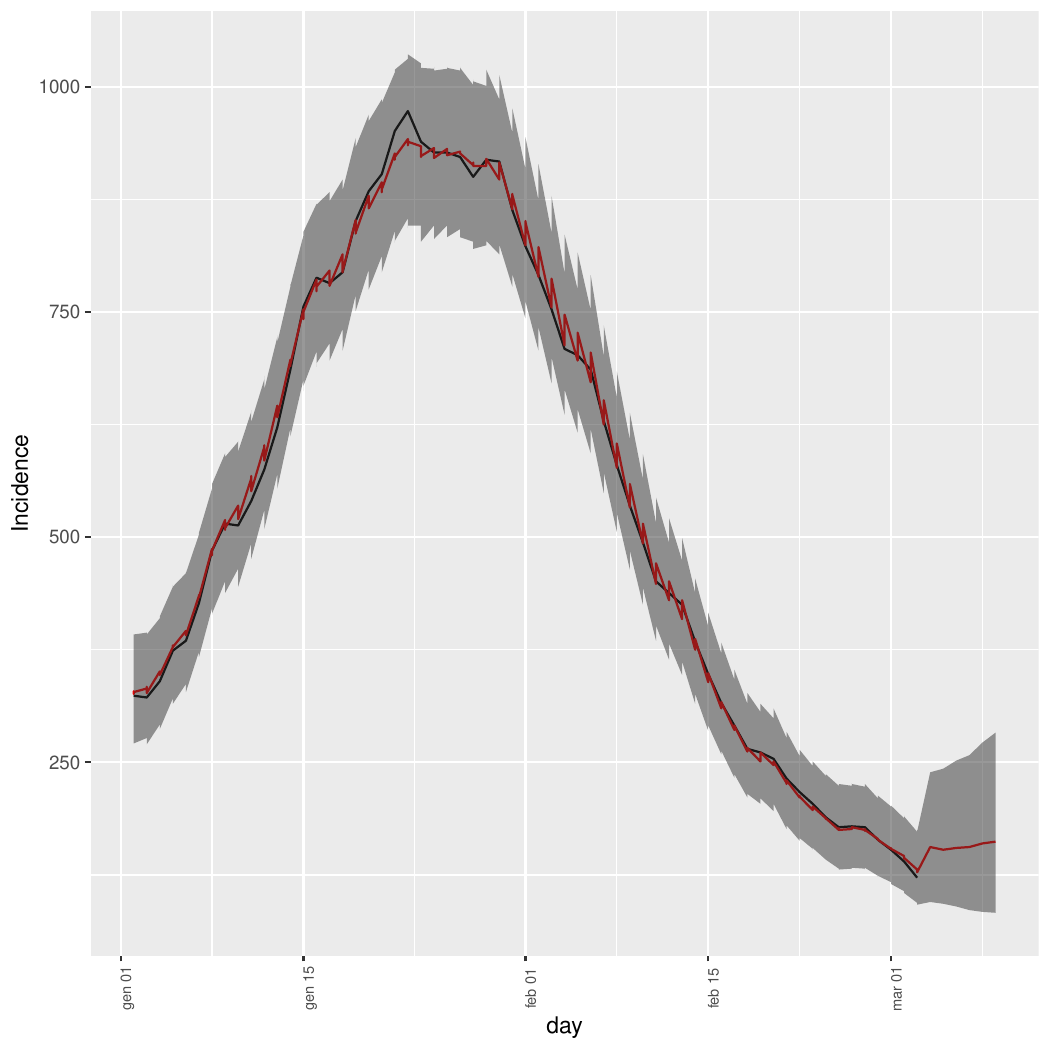}
        \caption{Daily incidence for year 2021 for the overall Spain. Posterior predicted distribution summarized by the median (red) and 99\% credible intervals (shaded grey area) for observed days (black) This is the estimation of \eqref{eqpp} for observed and future predictions starting from $d=1$ day ahead up to $d=7$ days ahead.}\label{fig.bayespred-spain}
    \end{center}
\end{figure}

From regional predictions is possible to obtain that for Spain by averaging over regional predictions and waiting according to their population. This is shown in Figure \ref{fig.bayespred-spain}.

From Figure \ref{fig.bayespred-spain} we can appreciate a general decrease of the incidence which is expected to be hardly below 100 cases per 100 thousand residents for the next week.

To further validate the model, the LSTM expert was trained up to the last observed 7 days, such that last 7 days have been predicted using only \eqref{eqpriorpred}.

\begin{figure}[H]
    \begin{center}
        \includegraphics[width=10cm]{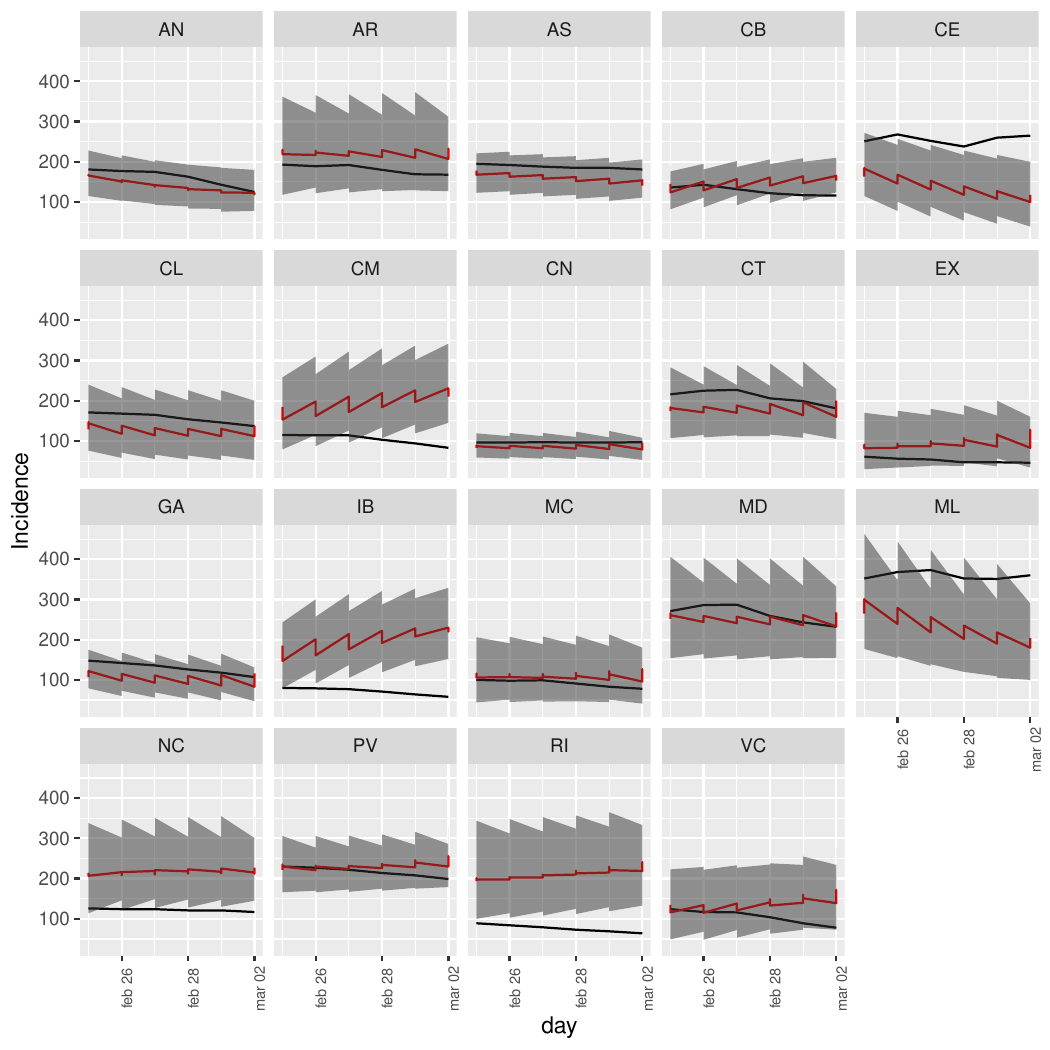}
        \caption{Predicted incidence $\widehat{y}_{ts}$ (red) with their 99\% CI (shadded grey area) and observed daily incidence (black) over last observed week when these observations have been never used to train the model.}\label{fig.lstmpredtest}
    \end{center}
\end{figure}

Figure \ref{fig.lstmpredtest} shows that the observed incidence is almost compatible taking into account the uncertainty provided by the Bayesian model.

Figure \ref{fig.lstmpredtest} further shows that the observed incidence in some regions is better to forecast than in others.

\subsection{Impact scenarios}\label{secscenarios}

This section is devoted to analyzing the impact of hypothetical scenarios and thus illustrates how the model can be eventually used by public policymakers. This section has also twofold purposes: firstly, to analyze the impact of plausible scenarios making predictions and secondly to attempt to interpret the overall model given that the LSTM projections lack a straightforward interpretation. For instance, the effect of a perturbation in a region and the effect on other regions can shed light on the spatial relation between regions that the LSTM learned from the data.

\subsubsection{Scenario 1: the case of increase in Madrid}

Suppose that during the last 20 days up to the last observed day, the number of deaths in the Madrid region was artificially incremented by 5\% every day from the day before. What should be the impact on other regions?

\begin{figure}[H]
    \begin{center}
        \includegraphics[width=10cm]{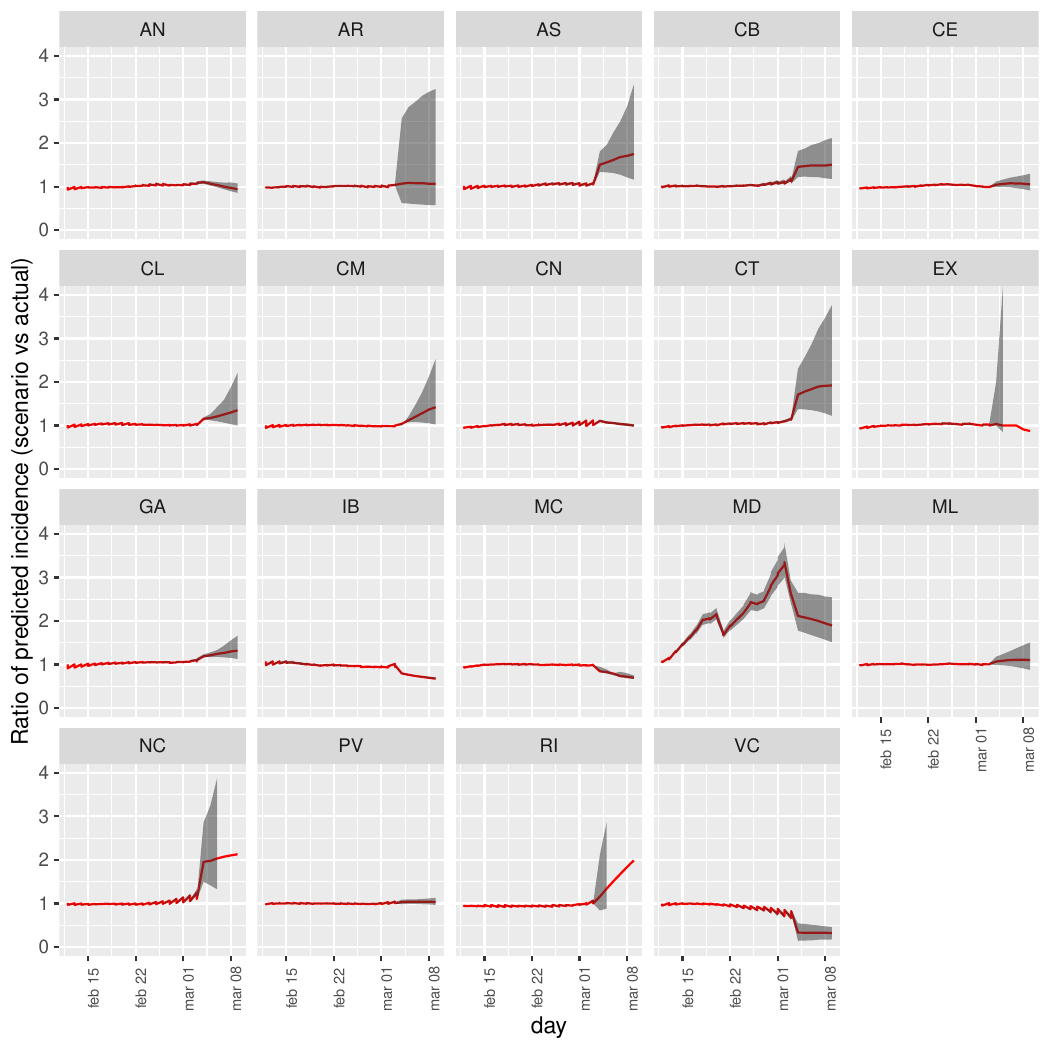}
        \caption{Effect on regions for a sudden increase of cases in Madrid. Ratio between the predicted incidence for the scenario against the actual predicted incidence. Red is the ratio among medians along with 99\% credible intervals (shaded grey area).}\label{fig.scenario1spain}
    \end{center}
\end{figure}

Figure \ref{fig.scenario1spain} reports the ratio between the predicted incidence for the scenario and actual estimations.

This scenario induces an increase in the incidence in almost all other regions.

\subsubsection{Scenario 2: cases increase in the Canary Islands}

\begin{figure}[H]
    \begin{center}
        \includegraphics[width=10cm]{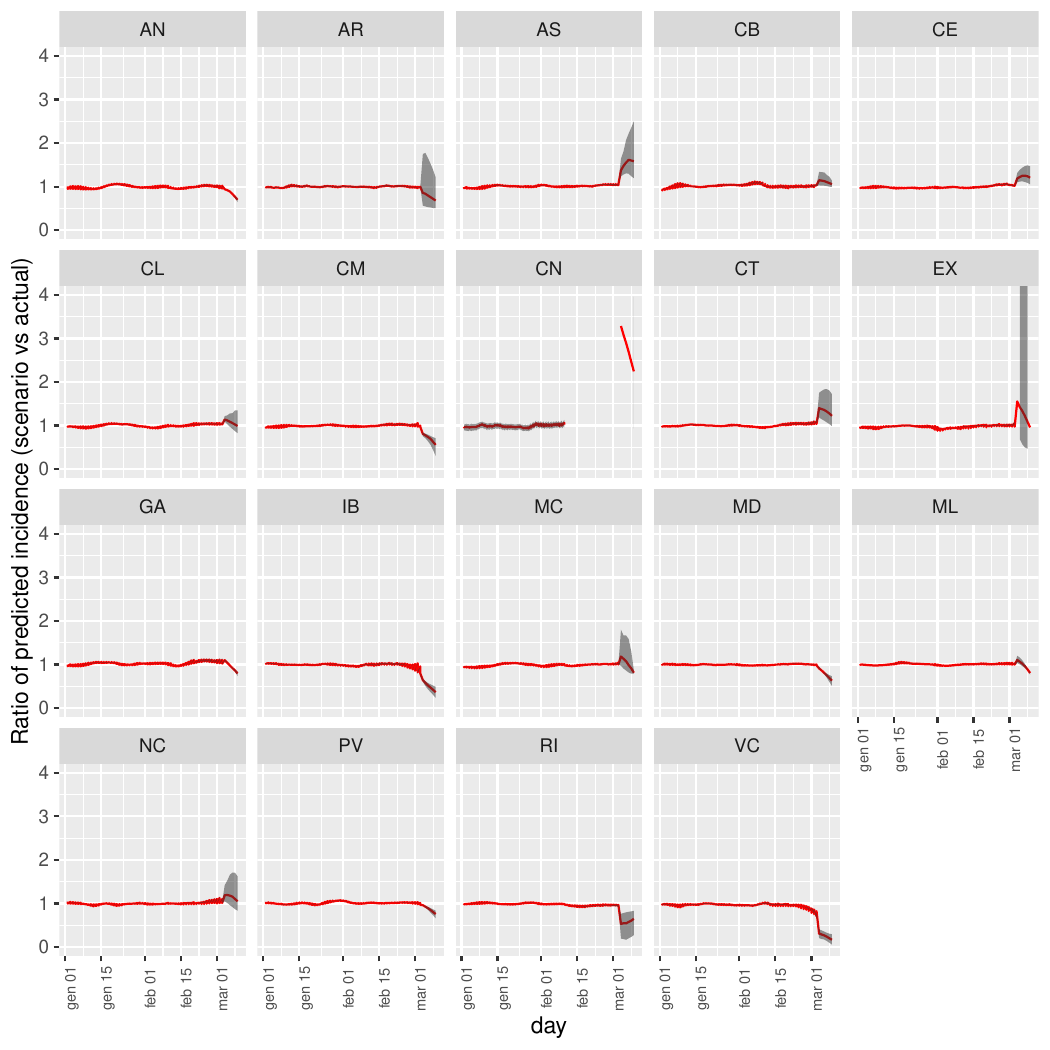}
        \caption{Effect on regions for a sudden increase of cases in Canary Islands. Ratio between the predicted incidence for the scenario against the actual predicted incidence. Red is the ratio among medians along with 99\% credible intervals (shaded grey area).}\label{fig.scenario2spain}
    \end{center}
\end{figure}

Suppose that the same number of cases as in scenario 1 occur in the Canary Islands, which has less populated and more disconnected than the Madrid region. Figure \ref{fig.scenario2spain} reports impact on all Spain.

Contrary to the previous scenario, an increase in the incidence in the Canary Islands does not induce a general strong increment of the incidence in other regions. This could be explained by the fact that being isolated, the dynamic of Canary Island is not so related to the Covid-19 dynamics in the rest of Spain.

\section{Further generalizations and remarks}\label{secconclusion}

The proposed modelling approach, more than a cleaned and specific model, allows taking into account part of the complexity of Covid-19 evolution in Spain by summarizing it in a complicated model structure represented by the bidirectional LSTM network, whose output is further processed into a Bayesian Poisson-Gamma model. This finally accounts for the randomness given by the underlying estimated evolution.

All exposition here can be applied to other definitions of regions, for instance, province or hospital areas or even areal data coming from geo-localization of cases. It can include other covariates on estimating Covid-19 evolution such as meteorological data if they were relevant or the presence of social restriction measures. The available commented code at \url{https://scabras.github.io/covid19-bayes-dl/} makes easy all this kind of generalization of the simplified analysis here proposed. This just assumes that all evolution process can be known from the observed counts and not from other sources of information as, for instance, the mobility.

The model here proposes incorporates temporal effects. The analysis could in principle be extended to temporal and spatial effects by using other DL architectures, for instance, Convolutional Neural Networks (CNN). This is a specific class of DL models suitable for analyzing structured samples as images \citep{r1-cnn} which can be generalized to time and spatial dimension. The spatial dimension is here not imposed into the model because of the reason explained in Section \ref{secintro} as it is not clear how to define the surrounding areas relevant for explaining Covid-19 evolution. It is well known that Covid-19 has been originally jointly detected in different areas of the world almost at the same time. We think it is better to capture the relations among the dynamics of different areas rather than imposition spatial effects among areas (especially for studies at big scales).

The important message here is that, with such a complex problem as the estimation of Covid-19 evolution, it may be not convenient to rely on a single one oriented approach (only machine learning or only parametric Bayes), but the combination of some would be more profitable. The drawback of this hybrid approach is that it becomes difficult to assess the overall theoretical reliability. For instance, we know very well that the Poisson-Gamma model leads to estimator which are closed to the true one (in mean squared error) when the prior precision is not too high, However, we don't have explicitly theoretical results on the consistency of LSTM models although there exists much successful application of them \cite{karpatne2017theory}. Even more, there are not theoretical consistency results when these two approaches are posed in sequence.

In contrast, we face a real (not theoretical) problem of predicting and forecasting Covid-19 evolution and further theory may come in the future to support or disprove the here proposed two-stage approach.

\section*{Acknowledgments}\label{secAcknowledgments}
The author is responsible for the content of this work, however, it is worth mentioning that this work would have been not possible without the encouragement of colleagues at the Department of Statistics of Universidad Carlos III de Madrid (Spain). The author is supported by the MINECO-Spain project PID2019-104790GB-I00.

\appendix

\section{Appendix section}\label{app}

\begin{figure}[H]
    \begin{center}
        \includegraphics[width=10cm]{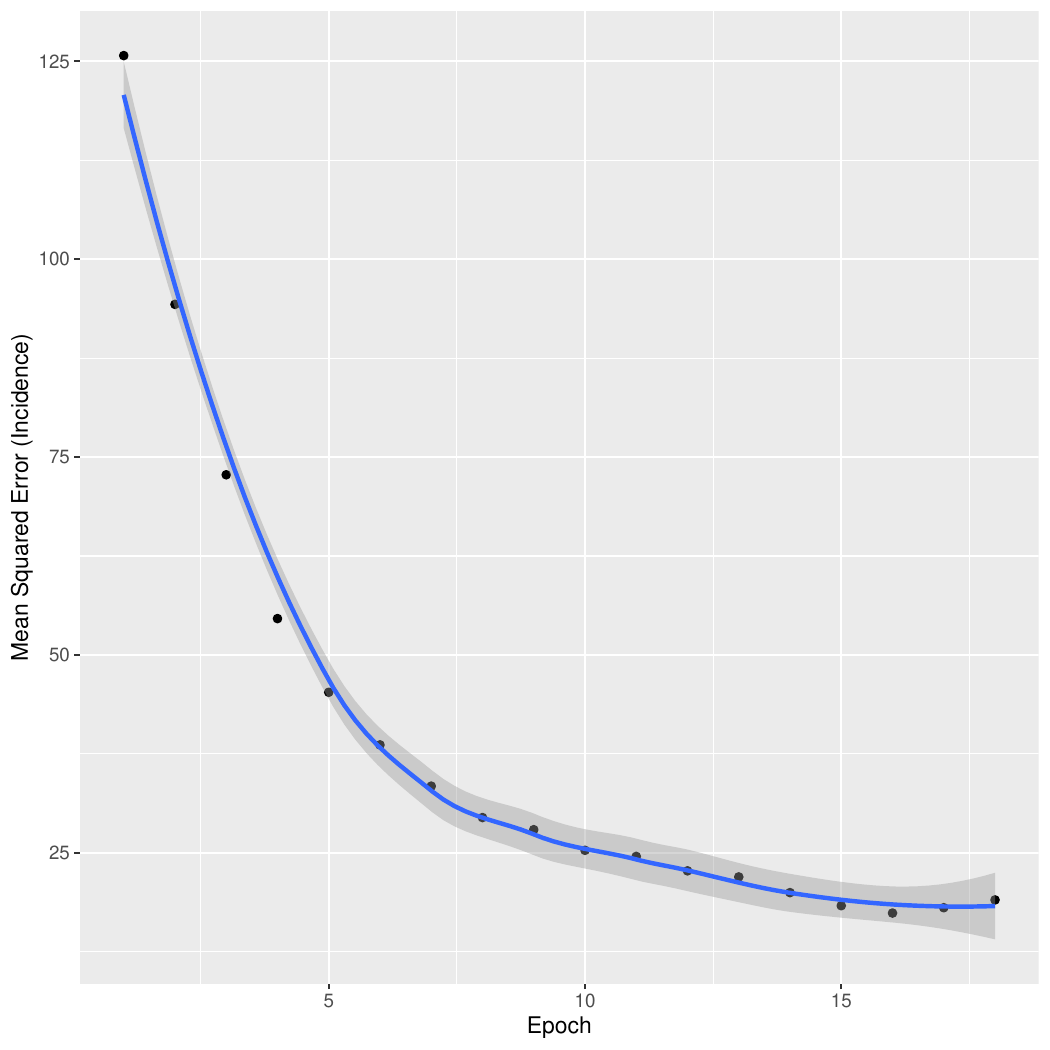}
        \caption{The mean squared error for counts is reported in vertical axis at each optimization step (horizontal axis) for train sequences.}\label{fig.history}
    \end{center}
\end{figure}

\begin{figure}[H]
    \begin{center}
        \includegraphics[width=10cm]{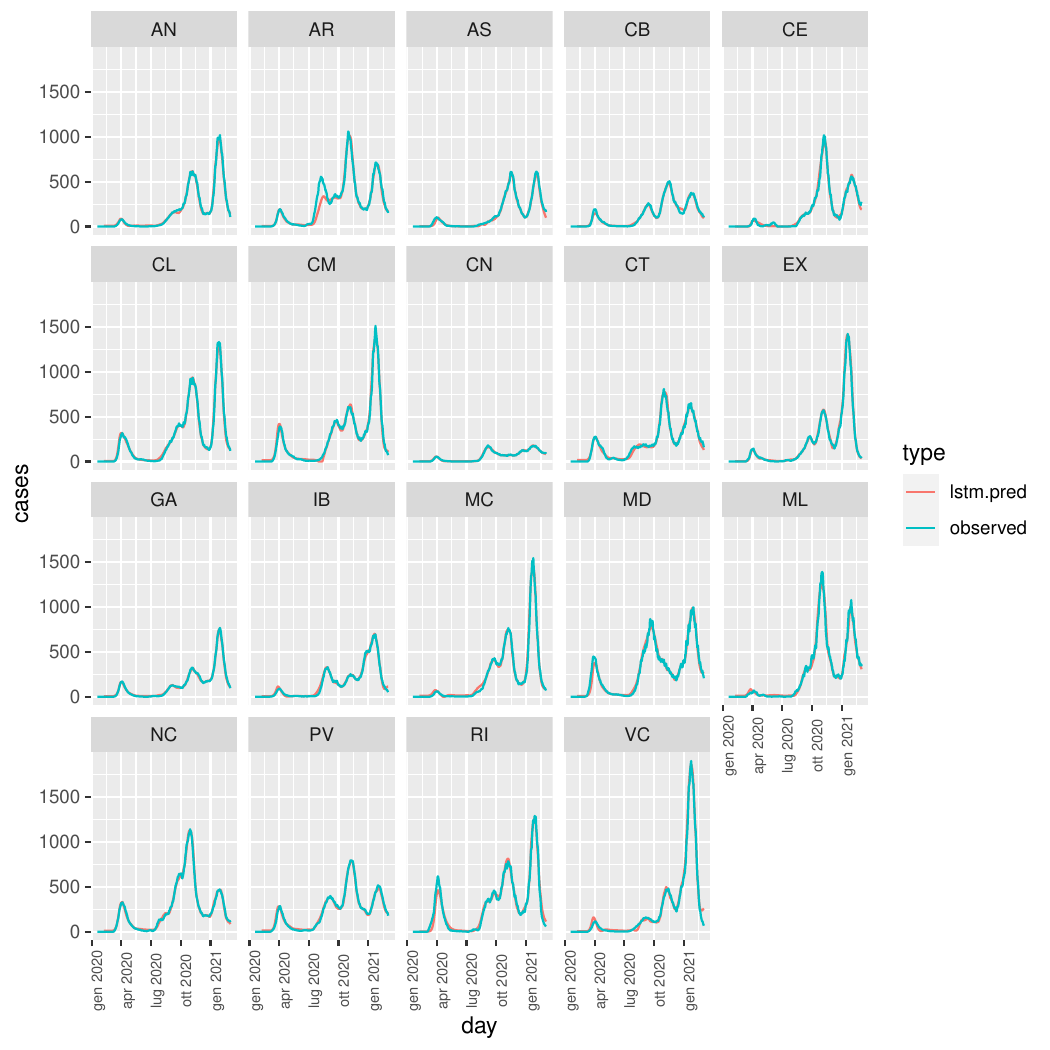}
        \caption{Predicted $\widehat{y}_{ts}$ and observed daily incidence over whole observed period.}\label{fig.lstmpredall}
    \end{center}
\end{figure}

\bibliographystyle{chicago}
\bibliography{bibliocovid19}

\end{document}